%% file: main2.tex
\documentclass[aps, prx, reprint, superscriptaddress, 10pt]{revtex4-2}
\usepackage{amsmath,amsfonts,amsthm,amssymb}
\usepackage{graphicx}
\usepackage[caption=false]{subfig}
\usepackage{placeins}

\usepackage{mathrsfs}
\newcommand{\blockcomment}[1]{}

\usepackage{etoolbox}
\patchcmd{\subsubsection}{\itshape}{\itshape \bfseries}{}{}

\begin{document}
	
\title{Thermodynamic Approach to Momentum Transport in Dense Fluids}
\author{Christopher Devik Fjeldstad}
\email{christopher.fjeldstad@ntnu.no}
\affiliation{Department of Mechanical and Industrial Engineering, Norwegian University of Science and Technology, NO-7491 Trondheim, Norway}
\author{Jonas Bueie}
\email{jonas.bueie@ntnu.no}
\affiliation{PoreLab, Department of Chemistry and Biomedical Science, Norwegian University of Science and Technology, NO-7491 Trondheim, Norway}
\author{Astrid S. de Wijn}
\email{contact author: astrid.dewijn@ntnu.no}
\affiliation{Department of Mechanical and Industrial Engineering, Norwegian University of Science and Technology, NO-7491 Trondheim, Norway}

\begin{abstract}
We present a new framework for extending Chapman-Enskog theory beyond the hard-sphere fluid model. Rather than relying on effective hard-sphere diameters, the approach makes use of on an exchange function which can be related to the thermodynamic properties of the system. We show that two existing extensions, including modified Enskog theory (MET), fit into this new framework. Based on our approach, we propose an alternative to MET that takes into account the potential interaction energy associated with the inter-particle interactions in the fluid. The proposed expression is applied to predicting the shear viscosity of several different simulated fluid models across a wide set of densities $0.05 \leq \rho^* \leq 0.8$ and temperatures $1.5 \leq T^* \leq 4.0$ in Lennard-Jones units.
The fluid models considered include both the Weeks-Chandler-Andersen (WCA) fluid and the Lennard-Jones (LJ) fluid. At low and intermediate density, here taken to be $\rho^* \leq 0.3$, we report mean relative prediction errors between $2\%$ and $4\%$ for both of these. Across all densities considered, the largest mean relative errors reported are $4.4\%$ and $8.1\%$ for the WCA fluid and LJ fluid respectively. We also investigate other interaction models, including a diatomic molecular model, in order to better understand the limitations of our approach.
\end{abstract}

\maketitle

\input{Sections_new_structure/Introduction}
\input{Sections_new_structure/Theoretical_Background}
\input{Sections_new_structure/Our_proposal}
\input{Sections_new_structure/Models_and_Methods}

\input{Sections_new_structure/Results}
\input{Sections_new_structure/Conclusion}

\acknowledgments
This work has been supported by the Research Council of Norway through its Centers of Excellence funding scheme, project number 262644, and FRIPRO project number 275507, as well as the National Infrastructure for Computational Science in Norway (UNINETT Sigma2), project number NN9573K.  ASdW gratefully acknowledges the hospitality of the Nordic Institute for Theoretical Physics (Nordita) and the Nordita Corresponding Fellowship.

\appendix
\input{Appendix/AppendixA}

\input{Appendix/AppendixB}

\bibliography{Bibliography.bib}

\end{document}

%% file: Sections_new_structure/Introduction.tex
\section{Introduction}
\label{sec:Introduction}

The transport properties of fluids are crucial for our understanding of many processes, both natural and industrial.
Non-equilibrium systems are challenging in general, because, unlike for equilibrium systems, there exists no general statistical mechanics theory yet for out-of-equilibrium systems.
For dealing with transport properties of real systems, we are thus often limited to ad-hoc approaches.
Kinetic theory for fluids 
and the development by Chapman and Enskog and others of approximate solutions to the Boltzmann transport equation in the 20th century have provided a framework for quantitative predictions of transport coefficients of fluids~\cite{ChapmanCowling,KirkwoodI}. 
However, due to the drastic approximations needed to make them tractable, kinetic approaches struggle when applied to real fluids and their predictive power is consequently limited. 
As such, when using kinetic-theory based approaches for quantitative predictions, engineers are forced to combine this with heavily ad-hoc and empirical approaches~(see, for example Refs.~\cite{Dymond1987, Yusibani2017}).
Thus, substantial effort in applied research has been aimed at alleviating the limitations of Chapman-Enskog theory, devising different approaches when applying it to increasingly complex fluids~\cite{KirkwoodIII, deWijn2008, UMLA2012, UMLA2014, Robles, Jervell2023}. Nevertheless, these approaches are still based around effective hard-sphere (HS) descriptions, and without detailed empirical input and numerous fitting parameters typically fail well below the critical density.
For a comparison of three distinct approaches, including the approach proposed by Kirkwood \textit{et al.}\cite{KirkwoodIII}, to calculating the shear viscosity of LJ fluids, see for example~\cite{Dyer}.
In addition to theoretical approaches, there has been extensive work done using simulations to study the transport properties of soft systems. The Lennard-Jones fluid has received a lot of attention. See for example~\cite{Meier}.

In this paper, we instead propose a general framework for calculating the density dependence of the transport coefficients based on the thermodynamic state of the system, and show that modified Enskog theory~\cite{Hanley1972} belongs to this class of general expressions. Rather than calculating effective hard-sphere properties such as the effective diameter, our approach turns the problem into one of finding an appropriate thermodynamic relation that contains information about the underlying exchange process. We specifically investigate a relation that takes into account the potential energy associated with the inter-particle interactions in the fluid.

We apply our resulting expression to three distinct types of fluid model systems that we simulate numerically for comparison.
They are 1) cut and shifted Mie fluids, including the common Weeks-Chandler-Andersen fluid~\cite{Weeks1971}, 2) full Mie fluids, including the Lennard-Jones fluid~\cite{Lennard-Jones1931}, and 3) a diatomic Lennard-Jones fluid. These systems are chosen as they allow us to study how different and increasingly complex interaction characteristics impact the accuracy of our approach. We simulate these fluids using molecular dynamics and obtain the viscosity numerically.

The remainder of this paper is structured as follows.
For a more thorough introduction into Chapman-Enskog theory, along with the most relevant expressions used to describe the thermodynamic properties of fluids, the reader is referred to Sec.~\ref{sec:Background}. The new framework is derived in Sec.~\ref{sec:Theory}. Model and simulation details can be found in Sec.~\ref{sec:method}. This Section also includes specific details on how the predicted shear viscosity is calculated for the various model systems under consideration. Lastly, the results of the numerical comparison can be found in Sec.~\ref{sec:Results}.

%% file: Sections_new_structure/Theoretical_Background.tex
\section{Theoretical Foundation}
\label{sec:Background}

Analytical approaches to calculating the transport properties of real dense fluids, including the approach being proposed in this work, are typically based on Chapman-Enskog theory, which links the transport properties of simple fluids to the radial distribution function (RDF) $g(r)$ of the fluid in equilibrium. 
We here provide a brief introduction to the most relevant expressions used to describe the thermodynamic and kinetic properties of simple fluid models, including the hard-sphere fluid model which commonly serves as a useful starting point for deriving analytical results~\cite{Mulero2008, HansenMcDonald2006}.

All fluids considered in the present work consist of $N$ identical particles confined to a volume $V$ so that the number density is given by $\rho = N/V$. The fluid is kept at a temperature $T$ and the resulting pressure is denoted by $p$. 

Before introducing Chapman-Enskog theory, we first go over some of the basic thermodynamics and equilibrium properties that will be needed.
The fluid compressibility factor is given by $Z = p\beta/\rho$, where $\beta = 1/k_\mathrm{B}T$ denotes the inverse temperature and $k_\mathrm{B}$ denotes the Boltzmann constant. 
Let the fluid model have only pairwise inter-particle interactions $\phi(r)$, where $r$ denotes the inter-particle distance.
The compressibility factor can be related to an integral over the RDF by means of the pressure equation~\cite{HansenMcDonald2006} 
\begin{equation}
\label{eq:Z-integral}
    Z - 1 = - \frac{2}{3}\pi\rho\beta\int^{\infty}_0\frac{d\phi(r)}{dr}g(r, \beta)r^3dr~.
\end{equation}
Analytical solutions to Eq.~(\ref{eq:Z-integral}) are not available for the general case.
However, for the hard-sphere fluid
Eq.~(\ref{eq:Z-integral}) takes the much simpler form~\cite{HansenMcDonald2006},
\begin{equation}
\label{eq:Z-HS}
    Z - 1 = V_\mathrm{excl} \rho g(\sigma^+_\mathrm{HS})~.
\end{equation}
Here $\sigma_\mathrm{HS}$ is the hard-sphere diameter, $V_\mathrm{excl} = 2\pi\sigma_\mathrm{HS}^3/3$ is the excluded volume of the hard spheres, and 
$g(\sigma^+_\mathrm{HS})$ is equal to $g(r)$ evaluated at point of contact, 
formally given by $\lim_{r \to \sigma^+_\mathrm{HS}} g(r)$, where the superscript denotes that the limit is taken from above.
The Carnahan-Starling equation of state~\cite{Carnahan1969} then provides a common and accurate approximation for $Z$ from which follows directly an expression for $g(\sigma^+_\mathrm{HS})$.

Now we return to the transport properties. These can be calculated by finding approximate solutions to the Boltzmann transport equation~\cite{Boltzmann}. Of special interest here is the shear viscosity $\eta$. In the low-density limit the shear viscosity $\eta_0 = \lim_{\rho \to 0} \eta$ of a weakly correlated fluid, such as hard-sphere fluids, is found to be~\cite{ChapmanCowling}
\begin{equation}
    \label{eq:eta_0}
    \eta_0 = \frac{5}{16\sigma_\mathrm{cross}^2\Omega^{*(2,2)}}\sqrt{\frac{mk_\mathrm{B}T}{\pi}}~,
\end{equation}
where $m$ is the hard-sphere mass, $\sigma_\mathrm{cross}$ is the cross-section diameter, and $\Omega^{*(2,2)}$ is the collision cross-section integral. 
For hard-sphere fluids $\sigma_\mathrm{cross} = \sigma_\mathrm{HS}$, and $\Omega^{*(2,2)} = 1$.

Although limited in their general applicability, hard-sphere models represent a crucial step in developing more generalized approaches to calculating transport properties of dense fluids.
Chapman-Enskog theory provides an approach to taking density into account, by considering the non-linear density dependence of the hard-sphere collision rate. The resulting expression for the shear viscosity of hard-sphere fluids is given by~\cite{ChapmanCowling}
\begin{equation}
\label{eq:Enskog}
\begin{split}
    \eta_\mathrm{Enskog} = \eta_0 V_\mathrm{excl} \rho  \bigg [ & \frac{1}{V_\mathrm{excl} \rho g(\sigma_\mathrm{HS}^+)} + 0.8 \\ 
                                                          & + 0.7614V_\mathrm{excl} \rho g(\sigma_\mathrm{HS}^+)  \bigg ]~,
\end{split}
\end{equation}
where $g(\sigma_\mathrm{HS}^+)$ shows up because it is related to the rate at which hard-sphere particles collide and therefore exchange momentum. 

Attempts to apply Chapman-Enskog theory to more complex fluid models
commonly involve finding effective hard sphere parameters for the fluid interaction. 
This includes the remarkably successful heuristic Enskog-$2\sigma$ model for simple liquids and gasses~\cite{UMLA2012, UMLA2014}, hard-sphere chain theory for simple hydrocarbons~\cite{deWijn2008}, and more analytical approaches such as that of Jervell and Wilhemsen~\cite{Jervell2023} which is based on revised Enskog theory~\cite{VanBeijeren1973, Lopez1983}.
However, these types of analytical approaches only provide reasonable predictions as long as the density remains low, i.e., well below the critical density of the fluid. 
In general, the use of effective hard-sphere diameters introduces severe limitations on the density-dependence of all involved quantities. As we know from experiments and simulations, complex density-dependence is common even for simple fluids.

The exception is modified Enskog theory (MET)~\cite{Hanley1972}, which does not use an effective hard sphere description. Based on the assumption that the thermal pressure can be used for calculations of an effective collision rate, MET provides an alternative route for predicting transport coefficients where all parameters in Chapman-Enskog theory can be related to experimentally accessible thermodynamic properties of the fluid. However, as with effective hard-sphere approaches, the accuracy of MET is greatly diminished at elevated density. It is understood that this is because MET, similar to effective hard sphere approaches, fails to take into account the impact of long-range interactions on the actual momentum exchange rate.

%% file: Sections_new_structure/Our_proposal.tex
\section{Our Proposed Framework}
\label{sec:Theory}

In this work, we aim to provide a more general framework based around an exchange function that does not necessarily depend on effective hard-sphere parameters.
In this section, we construct a general framework for our approach. We then show that MET actually belongs to this more general class of approaches.  The same is true for another extension, dipolar hard-sphere (DHS) kinetic theory~\cite{Pousaneh2020b}, used for predicting the viscosity of DHS fluids. Lastly, we propose a specific thermodynamic relation for the exchange function that takes into account the potential interaction energy associated with the inter-particle interactions, and motivate this choice.

\subsection{General framework}

We begin by noting that in Eq.~(\ref{eq:Enskog}) $g(\sigma_\mathrm{HS}^+)$ always appears together with $V_\mathrm{excl}\rho$.
We therefore introduce a convenient exchange function $\hat{g}$ that depends on the total excluded volume fraction $V_\mathrm{excl}\rho$, as well as pairwise correlations, which for hard spheres is given by
\begin{equation}
\label{eq:hat_g_definition}
    \hat{g} = V_\mathrm{excl} \rho g(\sigma_\mathrm{HS}^+)~.
\end{equation}
When applied to hard spheres, this quantity is directly proportional to the collision rate of hard spheres~\cite{HansenMcDonald2006}.
Making use of the fact that $V_\mathrm{excl}$ is independent of density, 
Eq.~(\ref{eq:Enskog}) can now be rewritten as
\begin{equation}
\label{eq:Enskog_modified}
    \frac{\eta_\mathrm{Enskog}}{\eta_0} = \lim_{\rho \to 0} \left [\frac{\hat{g}}{\rho} \right ] \rho \left [\frac{1}{\hat{g}} + 0.8 + 0.7614\hat{g} \right ]~,
\end{equation}
where substituting $\lim_{\rho \to 0} (\hat{g}/\rho)$ for $V_\mathrm{excl}$ ensures that the left hand side remains consistent with the intended low-density behavior, $\lim_{\rho \to 0} \eta_\mathrm{Enskog}/\eta_0 = 1$. Because $\lim_{\rho \to 0} g(\sigma_\mathrm{HS}^+) = 1$ for hard spheres, Eq.~(\ref{eq:hat_g_definition}) implies that $\lim_{\rho \to 0} ( \hat{g} / \rho) = V_\mathrm{excl}$ for hard-sphere fluids. The right hand side of Eq.~(\ref{eq:Enskog_modified}) captures all density dependent deviation from the low-density limit.

Attempts have been made to derive more accurate values for the coefficients in Eq.~(\ref{eq:Enskog_modified}), by taking into account higher-order correlations. See for example the work by Pieprzyk et al.~\cite{Pieprzyk}.
However, such modifications do not change the general structure of the Enskog expression, and thus do not affect the validity of our approach. Moreover, the corrections to the coefficients are very small compared to the effect of soft interactions.
We have verified that these small changes in the coefficients do not significantly affect our results.

Obtaining $\hat{g}$ for anything other than hard spheres is non-trivial. However, it does conveniently appear in Eq.~(\ref{eq:Z-HS}), the equation of state for hard spheres. In general, any equation of state has to reduce to Eq.~(\ref{eq:Z-HS}) in the limit of HS, which means that they must have the general form $Z - 1 = \hat{g} + X_\mathrm{res}$, where $X_\mathrm{res}$ is an appropriate residual function of the thermodynamic state, here defined as a property that vanishes for hard-sphere fluids. By reordering the general form it follows that
\begin{equation}
\label{eq:hat_g_from_Z}
    \hat{g} = Z - 1 - X_\mathrm{res}~.
\end{equation}
Combined with Eq.~(\ref{eq:hat_g_from_Z}), Eq.~(\ref{eq:Enskog_modified}) provides an expression that relates the shear viscosity of the fluid to properties of the thermodynamic state, rather than ill-defined effective hard-sphere parameters.
In this way, Eq.~(\ref{eq:Enskog_modified}) turns the problem of finding an expression for the shear viscosity into a problem of finding a suitable function for $X_\mathrm{res}$.

\subsection{Existing extensions}

MET and DHS kinetic theory are both examples of theories that can be expressed in terms of Eq.~(\ref{eq:Enskog_modified}).
In order to motivate a suitable choice of $X_\mathrm{res}$, it is useful to work this out in detail.

\subsubsection{Modified Enskog theory}
\label{sec:MET}
In MET, the true pressure $p$ in $Z$ in Eq.~(\ref{eq:Z-HS}), is replaced by the thermal pressure $p_\mathrm{T} = T(\partial p/ \partial T)_\rho$. This gives~\cite{Hanley1972}
\begin{equation}
    \label{eq:Z-MET}
    V^\mathrm{MET}_\mathrm{excl} \rho g(\tilde{\sigma}_\mathrm{HS}) = Z - 1 + T \left (\frac{\partial Z}{\partial T} \right )_\rho~,
\end{equation}
where $\tilde{\sigma}_\mathrm{HS}$ denotes an effective hard-sphere collision diameter.
By making use of the assumption that $\lim_{\rho \to 0}g(\tilde{\sigma}) = 1$, MET provides an expression for the excluded volume given by
\begin{equation}
    \label{eq:V_excl-MET}
    V^\mathrm{MET}_\mathrm{excl} = \lim_{\rho \to 0} \left [ \frac{Z - 1 + T \left (\frac{\partial Z}{\partial T} \right )_\rho}{\rho} \right ]~.
\end{equation}
Substituting Eq.~(\ref{eq:Z-MET}) and Eq.~(\ref{eq:V_excl-MET}) into Eq.~(\ref{eq:Enskog}), one arrives at the MET expression for the shear viscosity of a real fluid.
It follows straight forwardly that MET is equivalent to setting $X_\mathrm{res}$ to $X^\mathrm{MET}_\mathrm{res} = -T(\partial Z / \partial T)_\rho$ in Eq.~(\ref{eq:hat_g_from_Z}) and substituting the resulting expression for $\hat{g}$ into Eq.~(\ref{eq:Enskog_modified}).

The use of thermal pressure is justified theoretically in MET by an assumption that momentum exchange between particles is limited to effectively rigid collisions between the particle cores. The presence of any additional interactions will only impact the collision probability. 
This means that MET should only be expected to work at low density and high temperature.

\subsubsection{Dipolar hard-sphere theory}

Pousaneh and de Wijn~\cite{Pousaneh2020b} proposed DHS kinetic theory, derived for predicting the shear viscosity of DHS fluids with particles that consist of hard spheres with diameter $\sigma_\mathrm{HS}$ with a point dipole at their centre. This approach makes use of the fact that~\cite{Lee1989} for DHS fluids
\begin{equation}
    \label{eq:Z-DHS}
    \frac{2\pi}{3}\sigma_\mathrm{HS}^3 \rho g(\sigma_\mathrm{HS}^+) = Z - 1 - \beta u_\mathrm{pot}~,
\end{equation}
where $\beta u_\mathrm{pot} = \beta \langle U_\mathrm{pot}\rangle/N$ with $U_\mathrm{pot}$ denoting the total potential energy of the fluid. 
Equation~(\ref{eq:Z-DHS}) provides an expression that relates the true hard-core radial distribution function $g(\sigma^+_{HS})$, to $Z$ and $\beta u_\mathrm{ex}$. In DHS kinetic theory, the excluded volume of a DHS particle is taken simply as the excluded volume of the hard-sphere core $V^\mathrm{HS}_\mathrm{excl} = 2\pi\sigma^3_\mathrm{excl}/3$. By solving Eq.~(\ref{eq:Z-DHS}) for $g(\sigma_\mathrm{HS}^+)$ and inserting this expression, along with $V^\mathrm{HS}_\mathrm{excl}$, into Eq.~(\ref{eq:Enskog}), Pousaneh and de Wijn were able to predict the shear viscosity of DHS fluids obtained from molecular-dynamics simulations for a wide set of densities and strongly dipolar interactions with only two fit parameters: an effective dipole moment and an effective collision cross-section integral.

Predictions of the viscosity using DHS kinetic theory turn out to also fit into the framework based on Eq.~(\ref{eq:Enskog_modified}). Setting $X_\mathrm{res}$ to $X^\mathrm{DHS}_\mathrm{res} = \beta u_\mathrm{pot}$ in Eq.~(\ref{eq:hat_g_from_Z}) and substitution of the resulting expression for $\hat{g}$ into Eq.~(\ref{eq:Enskog_modified}) gives an expression for the shear viscosity of the DHS fluid that turns out to be mathematically equivalent to DHS kinetic theory. 
The only difference between the two approaches is semantic. It stems from the fact that, for DHS fluids, $\lim_{\rho \to 0} g(\sigma_\mathrm{HS}^+) > 1$, and so $\eta'_0$, the prime denoting the parameter as it appears in DHS kinetic theory, is not the true low-density viscosity. Rather $\eta'_0 = \eta_0\lim_{\rho \to 0} g(\sigma_\mathrm{HS}^+)$, or equivalently, $\eta'_0 = \eta_0\lim_{\rho \to 0} ({\hat{g}/\rho}) / V^\mathrm{HS}_\mathrm{excl}$.

\subsection{Choosing $\boldsymbol{X_\mathrm{res}}$}
\label{sec:X_res}

Having established what $X_\mathrm{res}$ looks like for two seemingly distinct theoretical approaches with different approximations, we now explore their similarities and differences. Based on this comparison, we propose a general choice for $X_\mathrm{res}$. 

The assumptions underlying MET are expected to hold at low density, and so a reasonable choice for $X_\mathrm{res}$ will have to behave the same to leading (linear) order in the density. 
If $X^\mathrm{MET}_\mathrm{res}$ is expanded in orders of $\rho$, the linear term, corresponding to the second virial approximation, is given by~\cite{Hanley1972}
\begin{equation}
\label{eq:X_res_1}
    X^\mathrm{MET}_\mathrm{res} \approx -T\frac{d B_2(T)}{d T}\rho~, 
\end{equation}
at low density, where $B_2(T)$ is the second virial coefficient. 

It turns out that DHS kinetic theory ($X^\mathrm{DHS}_\mathrm{res} = \beta u_\mathrm{pot}$) has the same leading order (linear) density term as MET.
To understand this, we first have to realize that $U_\mathrm{pot}$ is equal to the excess internal energy $U_\mathrm{ex}$ for DHS particles and any other particles that have no internal degrees of freedom that store potential energy.
The excess internal energy $U_\mathrm{ex}$, assuming no external fields, is defined in terms of only the spatial configurational contributions to the total partition function~\cite{HansenMcDonald2006}.
The two energies differ only when internal degrees of freedom that can store potential energy are present, e.g.\ bond stretching.
If expanded in orders of $\rho$, the first order term of $\beta u_\mathrm{ex} = \beta U_\mathrm{ex}/N$ is given by the right hand side of Eq.~(\ref{eq:X_res_1}). The reader is referred to Appendix~\ref{app:first_order_term} for a more thorough derivation of this result. 

The expression $X_\mathrm{res} = \beta u_\mathrm{ex}$ has a number of properties that make it an attractive choice.
It has the correct low-density behavior regardless of internal degrees of freedom.
We have also shown above that it reproduces the hard-sphere expression $\hat{g} = V_\mathrm{excl} \rho g(\sigma_\mathrm{HS}^+)$
for 
DHS fluids, and in Appendix~\ref{app:hat_g_exactness}, we also show that it holds for any spherically symmetric potential with a rigid hard-sphere core, and a long-range tail that decays $\propto 1/r^3$.
It is therefore worth exploring this choice for $X_\mathrm{res}$ for a broader set of fluid interaction models, although it seems likely that more suitable expressions exist for other systems.  For example, as we show in Appendix~\ref{app:hat_g_exactness}, when the tail decays more generally as $\propto 1/r^n$, the hard-sphere expression is reproduced when $X_\mathrm{res} = (n/3)\beta u_\mathrm{ex}$.

The implicit assumption that follows by setting $X_\mathrm{res} = \beta u_\mathrm{ex}$ can be explored by considering how $\beta u_\mathrm{ex}$ is related to $Z$ and $g(r)$.
For an isotropic and homogeneous fluid with spherically symmetric interactions $\beta u_\mathrm{ex}$ can be expressed as an integral over $g(r)$, giving~\cite{HansenMcDonald2006}
\begin{equation}
    \label{eq:u_ex-integral}
    \beta u_\mathrm{ex} = 2\pi\rho\beta\int^\infty_0 \phi(r)g(r)r^2dr~.
\end{equation}
Subtracting Eq.~(\ref{eq:u_ex-integral}) from Eq.~(\ref{eq:Z-integral}), reordering, and applying the product rule gives
\begin{equation}
    \label{eq:hat-g-integral}
    \begin{split}
    Z - & 1 - \beta u_\mathrm{ex} \\
        & = - \frac{2}{3}\pi\rho\beta \int^\infty_0 dr g(r) \frac{d\phi(r)r^3}{dr}~.
    \end{split}
\end{equation}
For hard spheres, which do not store internal interaction energy, the right-hand side is simply an average over inter-molecular forces, and reduces to the right-hand side of Eq.~(\ref{eq:Z-integral}).
For non-rigid particles, choosing $X_\mathrm{res} = \beta u_\mathrm{ex}$ is equivalent to averaging over the more complex derivative expression in the integrand.

Equation~(\ref{eq:hat-g-integral}) can be used to demonstrate that $X_\mathrm{res} = \beta u_\mathrm{ex}$ can not be a suitable choice in all cases. Consider for example what happens when $\phi(r) \propto 1/r^n$, as is the case for inverse power law fluids, and specifically the case when $n = 3$ for which Eq.~(\ref{eq:hat-g-integral}) reduces to $Z - 1 - \beta u_\mathrm{ex} = 0$. In general, one might expect $X_\mathrm{res} = \beta u_\mathrm{ex}$ to fail when interactions have extremely soft cores, i.e., when $n \to 3$. 
However, under such conditions, one should not expect Chapman-Enskog theory to perform well anyway.

%% file: Sections_new_structure/Models_and_Methods.tex
\section{Models and Methods}
\label{sec:method}

In order to test Eq.~(\ref{eq:Enskog_modified}) and the proposed relation $X_\mathrm{res} = \beta u_\mathrm{ex}$, we calculate the shear viscosity of 3 different fluid models and compare to results from numerical simulations.
We here describe and motivate the specific molecular models we consider and how we simulate them. We also provide a description of how we apply our approach to predicting the viscosity of these fluid models. 

The main focus of the proposed general framework is to link equilibrium properties to the density dependence of the transport coefficients. Since we mean to test this link, we do not want to introduce extra uncertainty stemming from inaccuracies in the equilibrium properties.  Therefore, some of the predictions made will rely on data extracted from simulations for these equilibrium properties when sufficiently accurate analytical expressions are not available. In the specific case of the LJ fluid, there are a multitude of available expressions for the free energy that are sufficiently accurate for our purpose, such as the one proposed by Thol \textit{et al.}~\cite{thol2016}. In general, the framework is only fully predictive for the density dependence when applied to fluids for which accurate expressions for the free energy already exist. Details on how such free energy expressions can be used can be found in Sec.~\ref{sec:calc_hat_g}.
Similarly, because we focus on the density dependence, rather than the low-density limit, we assume that knowledge of this limit is available by some other method. The details for how we calculate $\eta_0$ in the present work are given in Sec.~\ref{sec:eta_0_estimate}.

The models, simulation parameters, predictions, and numerical results will all rely on a dimensionless LJ unit system, i.e., in units of the particle mass $m$, characteristic interaction diameter $\sigma$, and characteristic energy scale $\epsilon$, as well as the Boltzmann constant $k_\mathrm{B}$. Reduced units are denoted as follows: mass $M^* = M/m$, distance $r^* = r/\sigma$, harmonic spring constant $k^* = k\sigma^2/\epsilon$, time $\tau^* = \tau(\epsilon / m \sigma^2)^{1/2}$, temperature $T^* = k_\mathrm{B}T/\epsilon$, number density $\rho^* = \rho \sigma^3$, and viscosity $\eta^* = \eta \sigma^3 / (\epsilon \tau)$.

\subsection{Interaction models}

We investigate 3 interaction models that have been chosen to test the validity and the key limitations of our proposed approach. We use these to explore the softness of the repulsive particle core, the long-range attractive interactions, and the presence of additional degrees of freedom.

To investigate the effect of repulsive and attractive interactions, we employ the common Mie potential~\cite{Mie1903}, a soft interaction potential often used to approximate real molecules.
The Lennard-Jones (LJ) interaction also belongs to this class.
The Mie potential is written as 
\begin{equation}
\label{eq:mie_potential}
    \phi^\mathrm{Mie} (r) = C \epsilon \left[ \left(\frac{\sigma}{r}\right)^{\gamma_\mathrm{rep}} - \left(\frac{\sigma}{r}\right)^{\gamma_\mathrm{att}} \right]~,
\end{equation}
where the exponents $\gamma_\mathrm{rep}$ and $\gamma_\mathrm{att}$ are interaction dependent parameters and $C = [\gamma_\mathrm{rep}/(\gamma_\mathrm{rep} - \gamma_\mathrm{att})][\gamma_\mathrm{rep}/\gamma_\mathrm{att}]^{\gamma_\mathrm{att} / (\gamma_\mathrm{rep} - \gamma_\mathrm{att})}$.
When $\gamma_\mathrm{rep} = 12$ and $\gamma_\mathrm{att} = 6$, we have the LJ potential. 

We consider first the purely repulsive case of cut and shifted Mie interactions, where the attractive part of the interaction is removed.
This represents a soft extension to the hard-sphere model for which molecular interactions are no longer instantaneous.
The cut and shifted Mie potential is truncated at the minima and shifted so that
\begin{equation}
\label{eq:wca-potential}
    \phi^\mathrm{Cut/Shifted} (r) = \begin{cases}  \phi^\mathrm{Mie} (r) + \epsilon, & r \leq \sigma \left( \frac{\gamma_\mathrm{rep}}{\gamma_\mathrm{att}} \right)^{\frac{1}{\gamma_\mathrm{rep} - \gamma_\mathrm{att}}} \\ 0, & r > \sigma \left( \frac{\gamma_\mathrm{rep}}{\gamma_\mathrm{att}} \right)^{\frac{1}{\gamma_\mathrm{rep} - \gamma_\mathrm{att}}} \end{cases}~.
\end{equation}

For the cut and shifted Mie fluids we will consider three parameter sets, which will allow us to probe the effects of the softness of the core: 1) The Weeks-Chandler-Anderson (WCA) potential $(\gamma_\mathrm{rep}, \gamma_\mathrm{att}) = (12, 6)$, 2) $(\gamma_\mathrm{rep}, \gamma_\mathrm{att}) = (10, 5)$, and 3) $(\gamma_\mathrm{rep}, \gamma_\mathrm{att}) = (8, 4)$. 

We also consider systems with full Mie interactions. This adds a long-range attractive component to the interaction and so momentum exchange is no longer limited to short-range interactions. 
When using the full Mie potential as stated in Eq.~(\ref{eq:mie_potential}), we will consider only two unique parameter sets: 1) the LJ potential, and 2) $(\gamma_\mathrm{rep}, \gamma_\mathrm{att}) = (10, 5)$. The parameter set $(\gamma_\mathrm{rep}, \gamma_\mathrm{att}) = (8, 4)$ without a cutoff is not included in our tests, as it presents computational challenges.  These are discussed in more detail in Sec.~\ref{sec:SimSetup}.

\begin{figure}
    \includegraphics[width=0.4\textwidth]{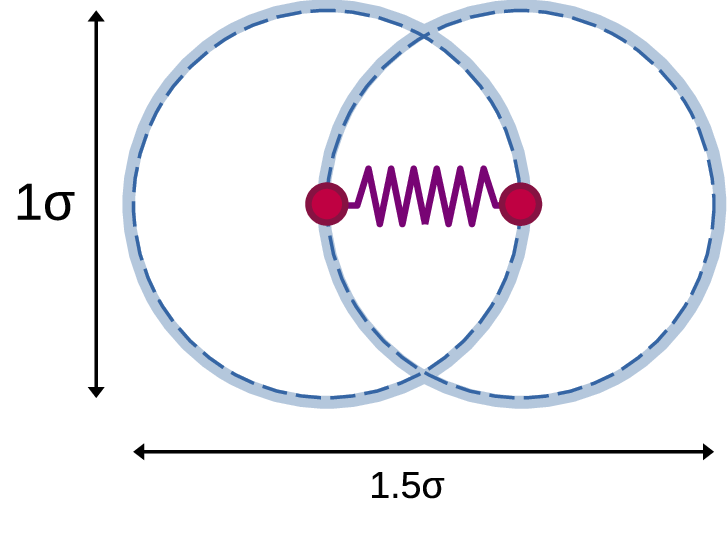}
    \caption{The geometry of diatomic molecules as modeled in this paper. The intra-molecular bond is modeled using a harmonic spring. Inter-particle interactions are modeled according to the Lennard-Jones potential.}
    \label{fig:molecule}
\end{figure}

Finally, we investigate a system of diatomic LJ molecules. The molecular structure introduces both rotational and vibrational degrees of freedom meaning that interactions are no longer fully elastic. In addition, potential energy is stored in bond stretching, making the distinction between $U_\mathrm{pot}$ and $U_\mathrm{ex}$ mentioned in Sec.~\ref{sec:X_res} relevant.
The diatomic LJ molecules consist of two Lennard-Jones particles, so that there are then $2N$ atoms in the system.
The interactions between the two atoms in the same molecule are modeled by a harmonic spring, while the interactions between atoms of different molecules are given by the LJ potential.
The diatomic molecule is illustrated in Fig.~\ref{fig:molecule}. The harmonic bond interaction energy is given by $\phi^\mathrm{Harmonic} = k(r_\mathrm{intra} - r_\mathrm{eq})^2/2$, where $r_\mathrm{intra}$ is the distance between the two atoms in the molecule, $r_\mathrm{eq} = 0.5\sigma$ is the equilibrium bond distance, and $k$ is the spring constant with units of energy/distance$^2$. By adjusting the spring constant we can probe the effect of changing the inelasticity of the microscopic interactions. We consider two cases: 1) a very rigid bond for which $k^* = 1000$ and 2) a slightly softer bond for which $k^* = 100$. 

\subsection{Simulation setup}
\label{sec:SimSetup}
The simulations of these fluids were performed using LAMMPS~\cite{Thompson2022}. The shear viscosity of the simulated fluids was obtained using the non-equilibrium method developed by M\"uller-Plathe~\cite{Muller-Plathe1999}. This approach relies on the non-physical swapping of momentum between particles in order to induce a velocity gradient ${\partial v_x}/{\partial z}$ without adding energy to the system. The resulting steady state system has a momentum flux $j_z$ along the gradient direction. The viscosity of the fluid is obtained from the momentum flux and velocity gradient in the steady state through the relation $j_z = - \eta {\partial v_x}/{\partial z}$.

The simulation box is rectangular, with the same size in the $x$ and $y$ directions for the base. In the $z$ direction the box is three times as long as in the $x$ and $y$ direction.  There are periodic boundary conditions in all directions.
Momentum is swapped along the $z$-axis, between particles in thin slabs at the edge and in the center of the region. Each slab has a width of $\mathrm{height}/100$. A pair is chosen for swapping once every $\delta \tau^* = 0.5$. This swapping rate is sufficiently small that the simulated fluids behave in a Newtonian manner.

Two checks have been performed to ensure Newtonian behavior. Firstly, it has been confirmed by visual inspection of the shear profile that the velocity gradient remains linear in all cases. Secondly, for a subset of parameters, we have rerun the simulations using both higher and lower swapping rates, confirming that this does not lead to any significant change in the estimated shear viscosity.

All simulations were run in the canonical ensemble (NVT), with a time step $\Delta\tau^* = 0.001$. The system is thermostatted using canonically sampled velocity re-scaling~\cite{Bussi2007}. This thermostat is chosen because it has minimal impact on the relevant dynamics of the system. 
Other more common thermostats, such as Nosé-Hoover and Langevin, have a more direct impact on the mechanics of the system, especially the ballistic part of the dynamics, and will therefore interfere more with the viscosity measurements.

Before starting the non-equilibrium simulation, the system is first equilibrated for a reduced time of 2000
with a thermostat damping time of $\tau^*_\mathrm{damp} = 100\Delta \tau^* = 0.1$.
For a subset of simulations, we have monitored the temperature and total potential energy during the equilibration period and concluded that 2000 is more than an order of magnitude more reduced time than is strictly needed for the fluid to reach equilibrium with the chosen thermostat settings. 

After equilibration, the M\"uller-Plathe algorithm is switched on, and the non-equilibrium simulation runs for an additional reduced time of 8000.
In order to further minimize the impact of the thermostat on the system dynamics, we switch to a longer thermostat damping reduced time of $\tau^*_\mathrm{damp} = 10000\Delta \tau^* = 10$.

Fluids using either the cut and shifted Mie potential or the full Mie potential to model particle interactions are simulated using $N = 10000$ particles at a reduced temperature $T^*$ in the range $1.5$ to $4.0$. 
However, because of phase separation in fluids with full Mie-(10, 5) interactions at $T^* = 1.5$, only the results for temperatures $T^* \geq 2.0$ are included in this specific case.

To speed up computation, for the full Mie potential a cutoff is applied to the potential tail end at distances $r^*_\mathrm{cut} = 4.0$ and $5.3$ for LJ fluids and Mie-(10, 5) fluids respectively. 
In order to properly capture the dynamics of a Mie fluid with interaction parameters $(\gamma_\mathrm{rep}, \gamma_\mathrm{att}) = (8, 4)$, a much larger cutoff would be needed to avoid finite-size effects, and thus also a much larger system would need to be simulated. This is computationally prohibitively expensive.

Diatomic LJ fluids are simulated using $N = 5000$ molecules, corresponding to $10000$ individual atoms. A cutoff is applied to the LJ potential tail end at distance $r^*_\mathrm{cut} = 4.0$.
The molecular fluids are simulated at three different reduced temperatures: $T^* = 3.0$, $3.5$, and $4.0$. We exclude temperatures below $3.0$, where we observe shear-induced phase separation.

Only densities $\rho^* \geq 0.05$ are considered for all fluid models. Below this, density related finite-size effects appear as the mean free path becomes comparable to our box size.

\subsection{Predicting the shear viscosity}
\label{sec:Predicting}

To test the effectiveness of Eq.~(\ref{eq:Enskog_modified}) combined with our form of $X_\mathrm{res}$ for predicting the viscosity of a fluid we need several parameters that are model dependent. 
This section lays out how the parameters are calculated in the present work.
This includes $\eta_0$ which determines the low-density limit of the resulting expression. 
Since the goal here is to test the approach, and not existing analytical expressions for these parameters, in many cases we minimize uncertainty due to imperfect analytical expressions by relying on simulation output data instead.

\subsubsection{Calculating $\boldsymbol{\eta_0}$}
\label{sec:eta_0_estimate}

Calculating $\eta_0$ using Eq.~(\ref{eq:eta_0}) generally requires extensive numerical analysis, due to the complexity of evaluating $\sigma_\mathrm{cross}^2\Omega^{*(2,2)}$. In the specific case of the LJ fluid, accurate and convenient empirical expressions for $\sigma_\mathrm{cross}^2\Omega^{*(2,2)}$ exist, such as the expression from Neufeld \textit{et al.}~\cite{Neufeld1972}. In this work, $\eta_0$ for the LJ fluid will be calculated by combining Eq.~(\ref{eq:eta_0}) and the Neufeld \textit{et al.} expression for $\sigma_\mathrm{cross}^2\Omega^{*(2,2)}$. In all other cases $\eta_0$ will be estimated based on simulation data, as detailed below.

We can obtain an initial estimate for $\eta_0$ by linear extrapolation based on two viscosity data points obtained from simulations at two different but low densities $\rho_1$ and $\rho_2$,
giving $\eta^\mathrm{linear}_0 = \eta_1 - \rho_1(\eta_2 - \eta_1)/(\rho_2 - \rho_1)$,
where $\eta_1$ and $\eta_2$ correspond to the viscosity of a fluid at densities $\rho_1$ and $\rho_2$.
We use $\rho_1 = 0.05$ and $\rho_2 = 0.1$.  We avoid densities below $0.05$, since our numerical simulations lose accuracy due to finite-size effects due to the long correlation length.
Since the second order is not insignificant in this density range, and typically has a positive coefficient,
linear interpolation will always underestimate $\eta_0$.
The true value must lie somewhere in between $\eta_1$ and $\eta^\mathrm{linear}_0$ and so we estimate the low-density limit from bisection, i.e.,
\begin{equation}
    \label{eq:estimating_eta_0_mean}
    \eta^\mathrm{estimate}_0 \approx \frac{\eta_1 + \eta^\mathrm{linear}_0}{2}~.
\end{equation}
The values obtained for $\eta_0$ are shown in Table~\ref{tab:eta_0_estimates}.
Although the accuracy of Eq.~(\ref{eq:estimating_eta_0_mean}) is likely somewhat limited, the exact value of $\eta_0$ has no impact on the analysis of our results, which is focused on the density dependence. A more accurate description of the low-density limit is therefore beyond the scope of this work.

\begin{table}
  \begin{tabular}{ r c c c c c c}
    \hline
    \hline
    \vspace{1px} \\
    \multicolumn{7}{c}{\textbf{Values for $\boldsymbol{\eta^*_0}$}} \\
    \vspace{1px} \\
    \hline
    \vspace{1px} \\
    \multicolumn{7}{c}{Cut and shifted Mie Potential} \\
    \hline
    $\boldsymbol{T^*}$ & \textbf{1.5} & \textbf{2.0} & \textbf{2.5} & \textbf{3.0} & \textbf{3.5} & \textbf{4.0} \\
    \hline
    $\boldsymbol{(\gamma_\mathrm{rep}, \gamma_\mathrm{att}) = (12, 6)}$ & 0.21 & 0.25 & 0.28 & 0.32 & 0.36 & 0.39 \\
    $\boldsymbol{(\gamma_\mathrm{rep}, \gamma_\mathrm{att}) = (10, 5)}$ & 0.21 & 0.25 & 0.29 & 0.33 & 0.36 & 0.39 \\
    $\boldsymbol{(\gamma_\mathrm{rep}, \gamma_\mathrm{att}) = (8, 4)}$ & 0.22 & 0.26 & 0.30 & 0.33 & 0.37 & 0.40 \\
    \hline
    \vspace{1px} \\
    \multicolumn{7}{c}{Full Mie Potential} \\
    \hline
    $\boldsymbol{T^*}$ & \textbf{1.5} & \textbf{2.0} & \textbf{2.5} & \textbf{3.0} & \textbf{3.5} & \textbf{4.0} \\
    \hline
    $\boldsymbol{(\gamma_\mathrm{rep}, \gamma_\mathrm{att}) = (12, 6)}$ & 0.16 & 0.21 & 0.25 & 0.29 & 0.33 & 0.36 \\
    $\boldsymbol{(\gamma_\mathrm{rep}, \gamma_\mathrm{att}) = (10, 5)}$ &      & 0.21 & 0.25 & 0.30 & 0.34 & 0.36 \\
    \hline
    \vspace{1px} \\
    \multicolumn{7}{c}{Diatomic Lennard-Jones Molecules} \\
    \hline
    $\boldsymbol{T^*}$ & \textbf{3.0} & \textbf{3.5} & \textbf{4.0} & & &\\
    \hline
    $\boldsymbol{k^* = 100}$ & 0.24 & 0.27 & 0.31 &&& \\
    $\boldsymbol{k^* = 1000}$ & 0.24 & 0.28 & 0.31 &&& \\
    \hline
    \hline
  \end{tabular}
  \caption{Values given for $\eta^*_0$ at various temperatures for the cut and shifted atomic Mie fluids (top), full atomic Mie fluids (middle), and diatomic LJ fluids (bottom), using the approach described in Sec.~\ref{sec:eta_0_estimate}. The values calculated for LJ fluids are given by combining the empirical expression for $\sigma_\mathrm{cross}^2\Omega^{*(2,2)}$, proposed by Neufeld \textit{et al.}~\cite{Neufeld1972}, with Eq.~(\ref{eq:eta_0}).
  All other values are estimated from simulation data using Eq.~(\ref{eq:estimating_eta_0_mean}).}
\label{tab:eta_0_estimates}
\end{table}

\subsubsection{Calculating $\boldsymbol{\hat{g}}$}
\label{sec:calc_hat_g}
To obtain $\hat{g}$ for LJ fluids, we make use of the excess Helmholtz free energy $F_\mathrm{ex}$.
There are many semi-empirical expressions for the free energy in the literature~\cite{Stephan2020}.
We can combine this with the thermodynamic relations 
$Z - 1 = \rho ( \partial \beta f_\mathrm{ex} / \partial \rho)_T$, and $\beta u_\mathrm{ex} = -T ( \partial (\beta f_\mathrm{ex}) / \partial T )_{\rho}$ where $f_\mathrm{ex} = F_\mathrm{ex}/N$ denotes the per-particle excess Helmholtz free energy.
We can then obtain $\hat{g}$ from our $X_\mathrm{res}= \beta u_\mathrm{ex}$ by inserting this into Eq.~(\ref{eq:hat_g_from_Z}), 
from which it follows that
\begin{equation}
\label{eq:hat_g-Helmholtz}
    \hat{g} = \rho \left ( \frac{\partial \beta f_\mathrm{ex}}{\partial \rho} \right)_T + T\left ( \frac{\partial (\beta f_\mathrm{ex})}{\partial T} \right )_{\rho}~.
\end{equation}
Similarly, for the sake of comparison, a prediction can be made using MET for which
\begin{equation}
\label{eq:hat_g_MET-Helmholtz}
    \hat{g}_\mathrm{MET} = \rho \left ( \frac{\partial \beta f_\mathrm{ex}}{\partial \rho} \right)_T + \rho T\left ( \frac{\partial ^2 (\beta f_\mathrm{ex})}{\partial T \partial \rho} \right )_{\rho}~.
\end{equation}
All predictions for the shear viscosity of a LJ fluid  presented in this paper have been calculated using an expression for $\hat{g}$ that is arrived at by combining either Eq.~(\ref{eq:hat_g-Helmholtz}) or~(\ref{eq:hat_g_MET-Helmholtz}) with the Thol \textit{et al.} expression for $f_\mathrm{ex}$~\cite{thol2016}. It has been checked that this provides the required accuracy for the range of parameters included in the present work.

In all other cases, there are no sufficiently accurate expressions available for the free energy.
Therefore, we calculate $\hat{g}$ using Eq.~(\ref{eq:hat_g_from_Z}) and thermodynamic output from equilibrium simulations.
The compressibility factor $Z = \beta p/\rho$ uses the calculated pressure of the simulated fluid~\cite{Thompson2009}. The excess internal energy is equal to the average simulation pairwise interaction energy. All predictions for the shear viscosity of model fluid systems using a cut and shifted Mie potential, the soft Mie-(10, 5) potential, or the diatomic LJ molecules, are calculated using equilibrium simulation output data.

\subsubsection{Calculating $\boldsymbol{\lim_{\rho \to 0} (\hat{g}/\rho)}$}
\label{sec:low_density}

For the monatomic fluids, we evaluate $\lim_{\rho \to 0}(\hat{g}/\rho)$ by expressing it in terms of a known interaction potential $\phi(r)$ for an isotropic homogeneous fluid by using that $\lim_{\rho \to 0} g(r) = e(r)$ where $e(r) = \exp{[-\beta \phi(r)]}$~\cite{HansenMcDonald2006}. Replacing the left hand side of Eq.~(\ref{eq:hat-g-integral}) with $\hat{g}$, substituting $g(r)$ for $e(r)$ on the right hand side, and dividing both sides by $\rho$, one arrives at 
\begin{equation}
    \label{eq:V_excl-integral}
    \lim_{\rho \to 0} \left [ \frac{\hat{g}}{\rho} \right ] = - \frac{2}{3}\pi\beta \int^\infty_0 dr \exp{[-\beta \phi(r)]} \frac{d\phi(r)r^3}{dr}~,
\end{equation}
For fluid models characterized by a single atomic Mie-type interaction, $\phi(r)$ is available and so $\lim_{\rho \to 0} \hat{g}/\rho$ can be calculated by numerical integration of Eq.~(\ref{eq:V_excl-integral}). 
Although derived for $X_\mathrm{res} = \beta u_\mathrm{ex}$, Eq.~(\ref{eq:V_excl-integral}) also applies to MET, due to the low density equivalence discussed in Sec.~\ref{sec:X_res}.

For diatomic molecules, $\phi(r)$ is not well-defined, and so we cannot use numerical integration of Eq.~(\ref{eq:V_excl-integral}). Instead, we estimate $\lim_{\rho \to 0} \hat{g}/\rho$ from equilibrium simulation data by linear extrapolation from two densities $\rho_1$ and $\rho_2$,
\begin{equation}
\label{eq:hat_g_lim_estimate}
    \lim_{\rho \to 0} \left [\frac{\hat{g}}{\rho} \right ] \approx  \frac{\hat{g}_1}{\rho_1} - \frac{(\hat{g}_2 / \rho_2) - (\hat{g}_1 / \rho_1)}{\rho_2 - \rho_1}\rho_1~,
\end{equation}
where $\hat{g}_1$ and $\hat{g}_2$ correspond to $\hat{g}$ obtained at densities $\rho_1$ and $\rho_2$.  We use $\rho_1 = 0.05$ and $\rho_2 = 0.1$. Tabulated values for $\lim_{\rho \to 0} (\hat{g}/\rho)$, as estimated for the diatomic Lennard-Jones fluids, can be found in Table~\ref{tab:V_excl_estimate}.

\begin{table}
\centering
\begin{tabular}{ p{0.3\linewidth} p{0.3\linewidth} p{0.3\linewidth} }
\hline
\hline
\vspace{1px} \\
\multicolumn{3}{c}{\textbf{Estimates for $\boldsymbol{\lim_{\rho \to 0} (\hat{g}/\rho)}$}}\\
\vspace{1px} \\
\hline
 & $k^* = 100$ & $k^* = 1000$ \\
\hline
$T^* = 3.0$ & 4.97 & 4.91 \\
$T^* = 3.5$ & 4.37 & 4.22 \\
$T^* = 4.0$ & 4.07 & 3.85 \\
\hline
\hline
\end{tabular}
\caption{Values for $\lim_{\rho \to 0} (\hat{g}/\rho)$ for diatomic LJ fluids, estimated using Eq.~(\ref{eq:hat_g_lim_estimate}).}
\label{tab:V_excl_estimate}
\end{table}

%% file: Sections_new_structure/Results.tex
\section{Results}
\label{sec:Results}

We here present and discuss the viscosity predictions made as outlined in Sec.~\ref{sec:Predicting} using $X_\mathrm{res} = \beta u_\mathrm{ex}$, and compare them to the results from numerical simulations. 

\subsection{The shear viscosity of atomic Mie-type fluids}
\begin{figure}
    \centering
    \subfloat[]{
        \includegraphics[width=0.45\textwidth]{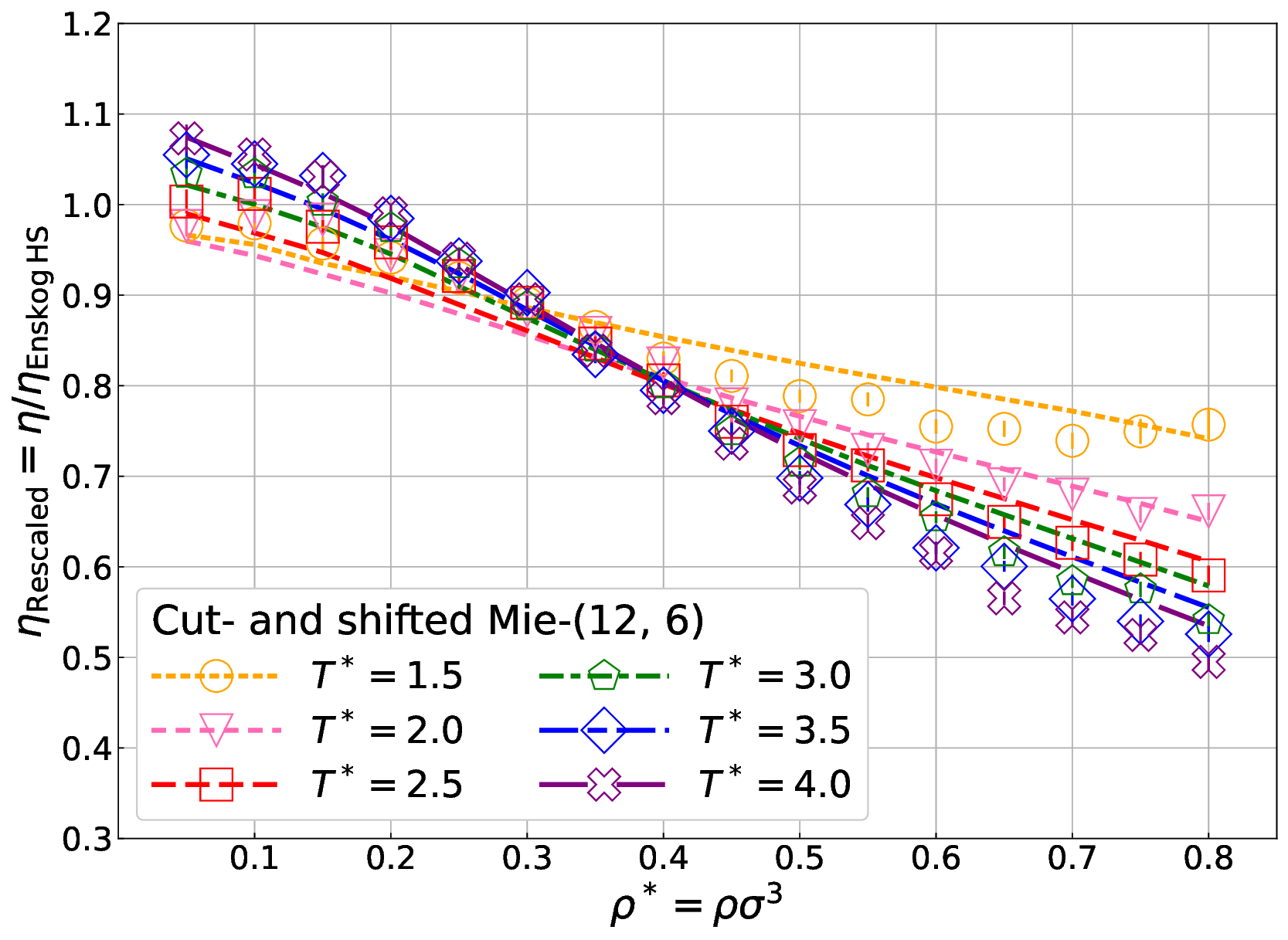}
        \label{subfig:WCA-12}}
        \newline
    \subfloat[]{
        \includegraphics[width=0.45\textwidth]{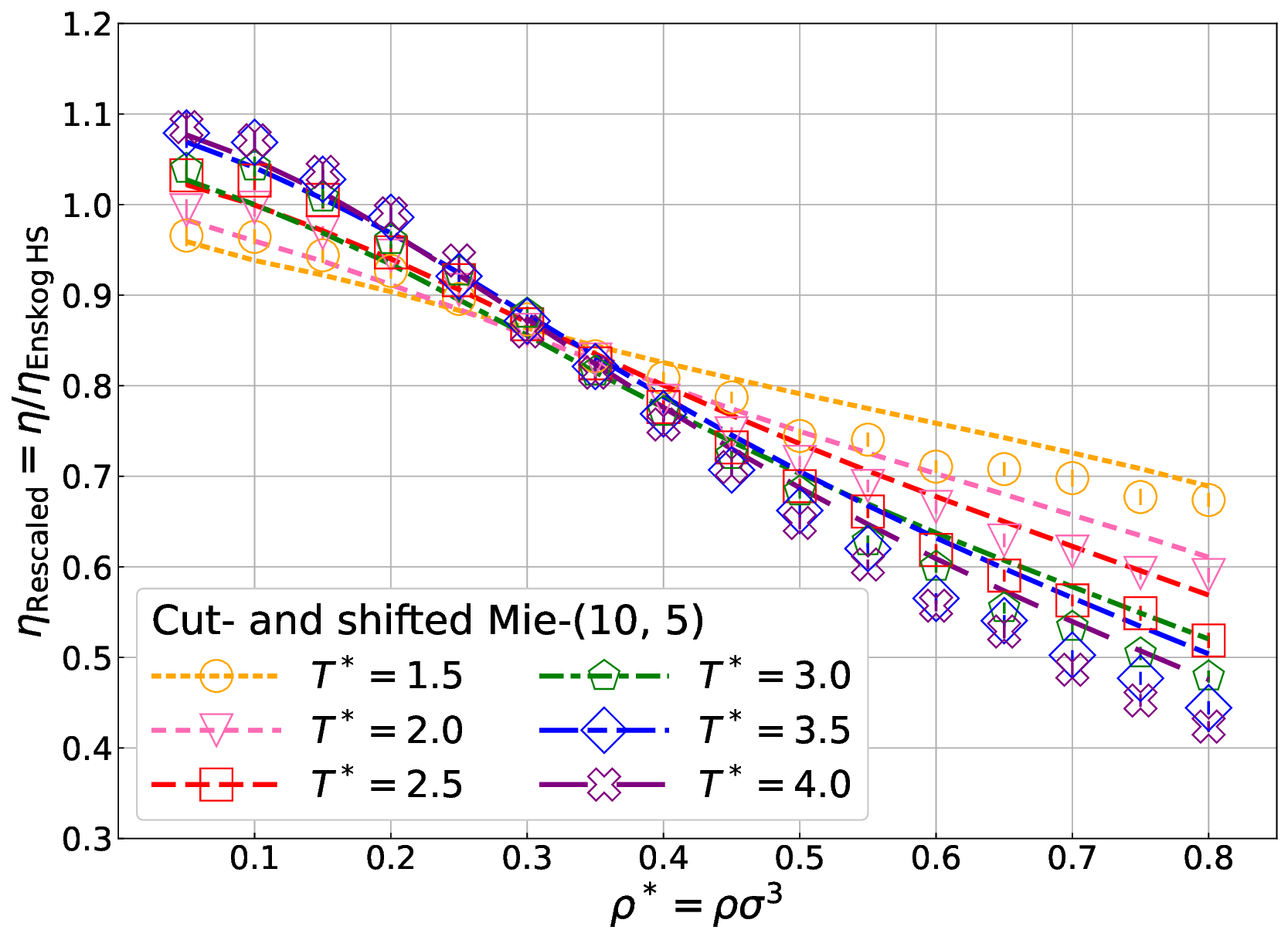}
        \label{subfig:WCA-10}}
        \newline
    \subfloat[]{
        \includegraphics[width=0.45\textwidth]{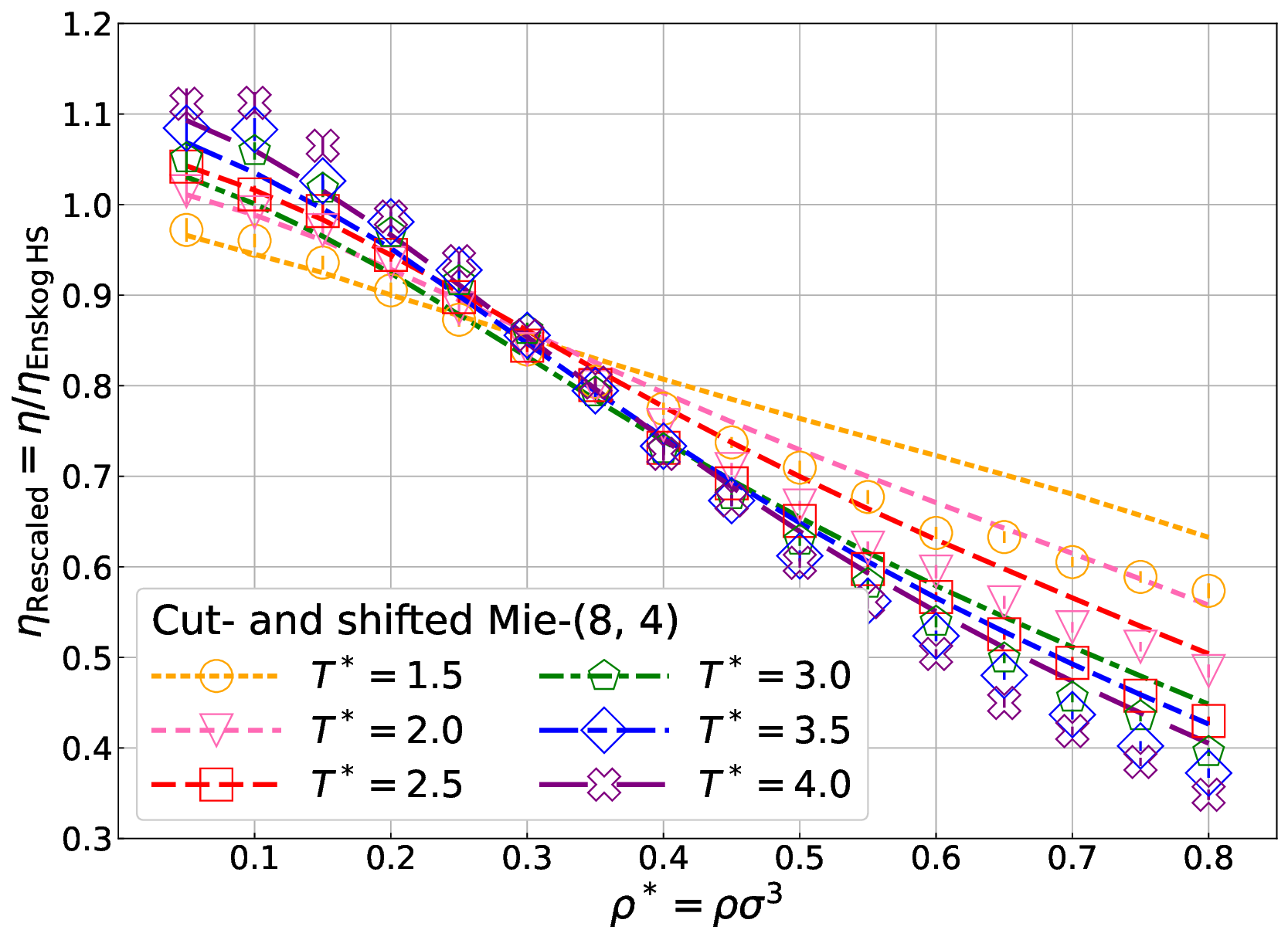}
        \label{subfig:WCA-8}}
    \caption{Rescaled shear viscosity from simulations and predictions for (a) WCA fluids, (b) cut and shifted Mie-(10, 5) fluids and (c) cut and shifted Mie-(8, 4) fluids, using $X_\mathrm{res} = \beta u_\mathrm{ex}$. Each temperature is assigned a symbol and corresponding dash style as denoted in the legend. The values used for $\eta_0$ are estimated using Eq.~(\ref{eq:estimating_eta_0_mean}), $\hat{g}$ is calculated based on equilibrium data from simulations as described in Sec.~\ref{sec:calc_hat_g}, and the low-density limit $\lim_{\rho \to 0} (\hat{g}/\rho)$ is calculated analytically using Eq.~(\ref{eq:V_excl-integral}). The shear viscosity is rescaled using the Enskog prediction for a hard-sphere fluid with diameter $\sigma_\mathrm{HS} = \sigma$ and mass $m = 1$.}
    \label{fig:viscosity_mie_cut}
\end{figure}

\begin{figure}
    \centering
    \subfloat[]{
        \includegraphics[width=0.45\textwidth]{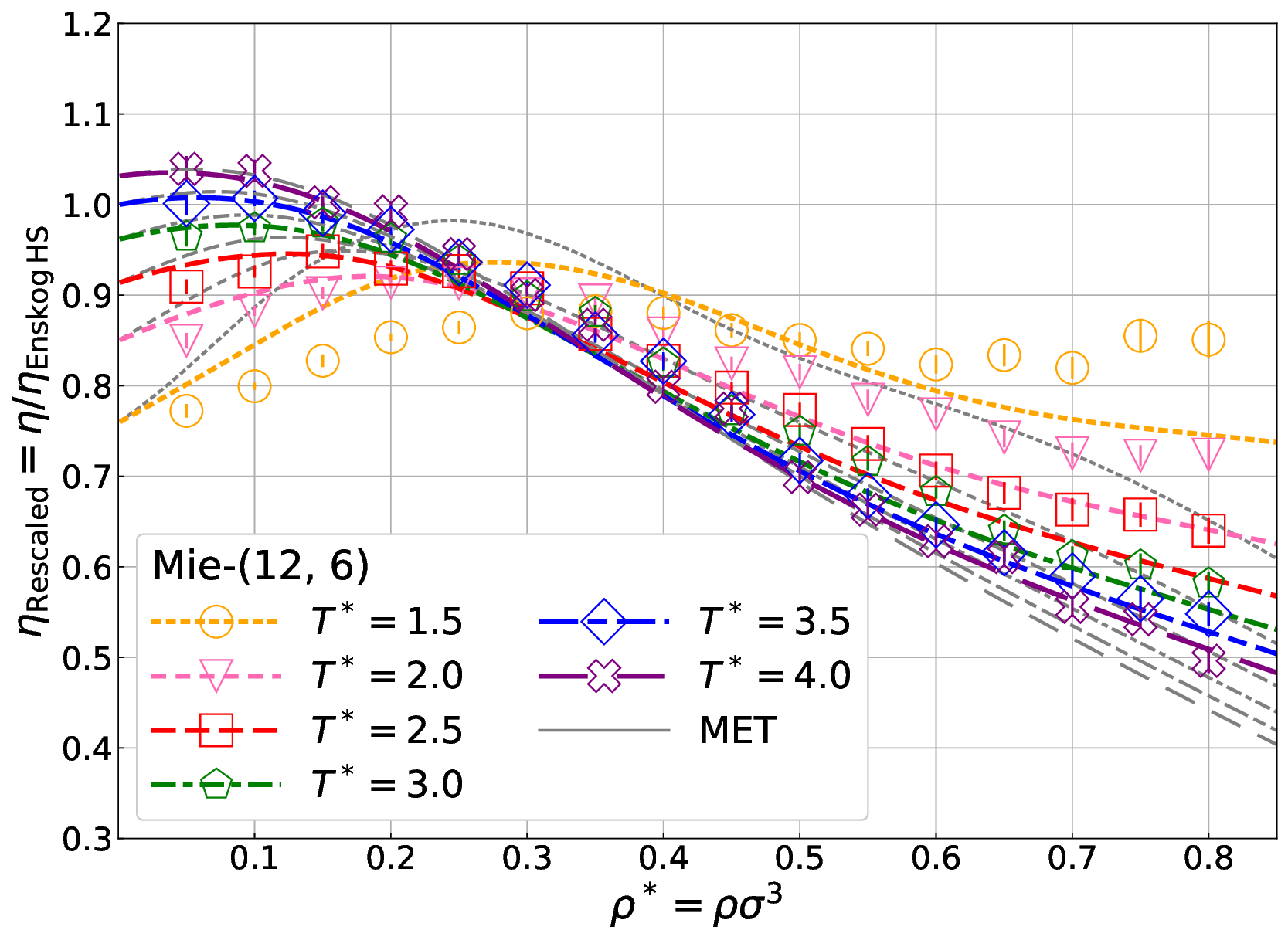}
        \label{subfig:LJ-12}}
        \newline
    \subfloat[]{
        \includegraphics[width=0.45\textwidth]{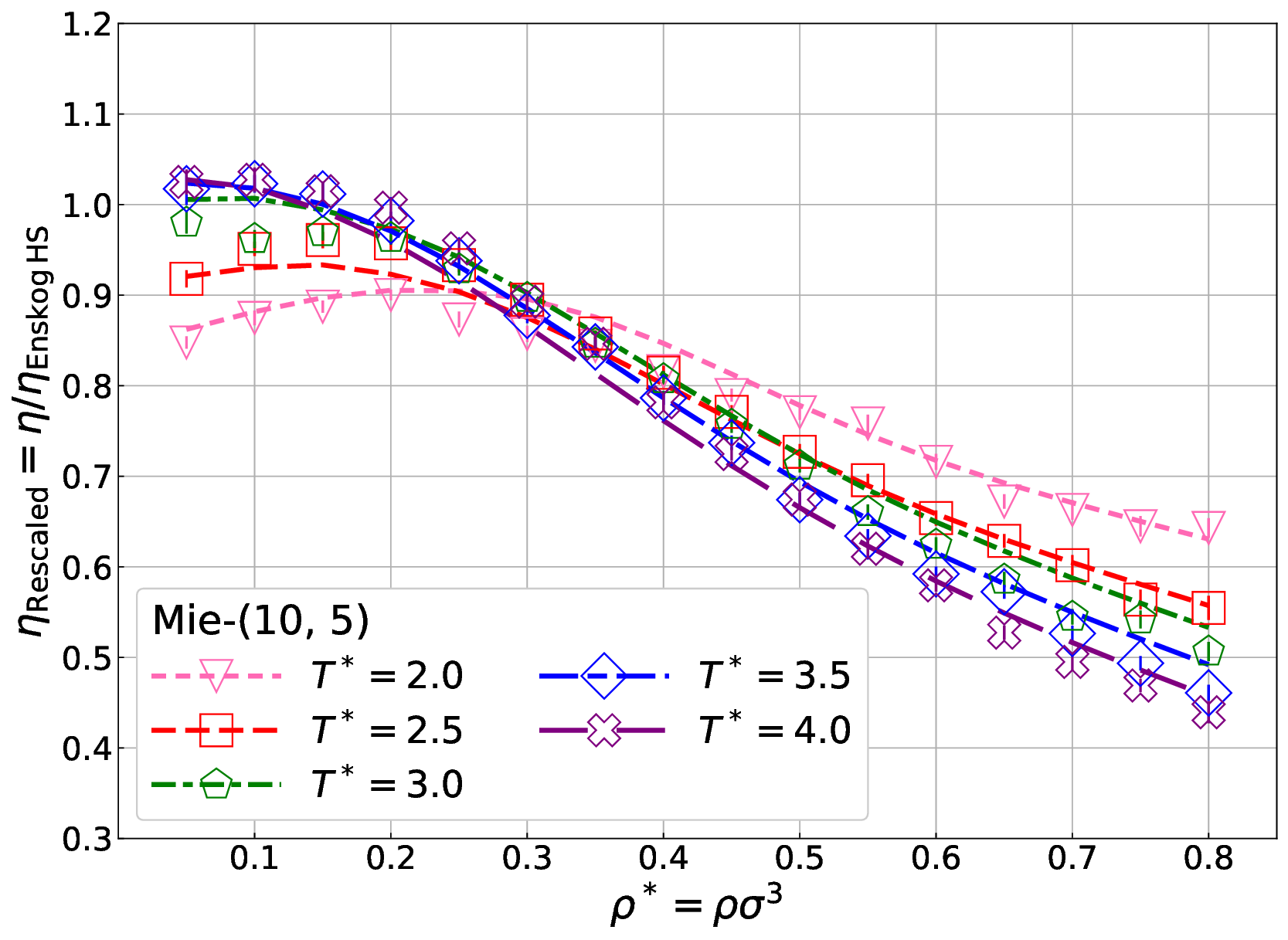}
        \label{subfig:LJ-10}}
    \caption{Rescaled shear viscosity from simulations and predictions for (a) LJ fluids and (b) Mie-(10, 5) fluids, using $X_\mathrm{res} = \beta u_\mathrm{ex}$. Each temperature is assigned a symbol and associated dash style as denoted in the legend. For comparison in Fig.\ (a) the light thin gray lines give the MET prediction with dash styles given by the corresponding temperature. 
    In (a), the values used for $\eta_0$ are calculated using Eq.(\ref{eq:eta_0}) along with the empirical expression given by Neufeld \textit{et al.}~\cite{Neufeld1972}, and $\hat{g}$ is calculated using Eq.~(\ref{eq:hat_g-Helmholtz}) along with the Thol \textit{et al.}~\cite{thol2016} expression for the free energy. Similarly, the MET prediction is calculated using Eq.~(\ref{eq:hat_g_MET-Helmholtz}).
    In (b) $\eta_0$ is estimated using Eq.~(\ref{eq:estimating_eta_0_mean}), and $\hat{g}$ is calculated based on equilibrium data from simulations as described in Sec.~\ref{sec:calc_hat_g}. In both (a) and (b), the low-density limit $\lim_{\rho \to 0} (\hat{g}/\rho)$ is calculated analytically using Eq.~(\ref{eq:V_excl-integral}).
    The shear viscosity is rescaled using the Enskog prediction for a hard-sphere fluid with diameter $\sigma_\mathrm{HS} = \sigma$ and mass $m = 1$.}
    \label{fig:viscosity_mie}
\end{figure}

Figures~\ref{fig:viscosity_mie_cut} and~\ref{fig:viscosity_mie} show a comparison of the predicted viscosity with data for the shear viscosity of cut and shifted Mie fluids and full Mie fluids respectively. For visual clarity, all viscosity data and corresponding predictions have been rescaled using the Enskog prediction for hard spheres given by Eq.~(\ref{eq:Enskog}) with $\sigma_\mathrm{HS} = \sigma$.

In all cases, the predicted shear viscosity is close to the numerical shear viscosity of the simulated model fluids, with typical relative errors $\delta_\mathrm{error}$ lying in the range $\delta_\mathrm{error} \lesssim 5\%$ when the density is low.
Even in the most extreme cases, $\delta_\mathrm{error}$ is less than~15\%.
When $\delta_\mathrm{error} > 5\%$, the system is typically at very high density, or close to the critical temperature $T_\mathrm{c}$ of the fluid.
An example is the prediction made for a LJ fluid at $T^* = 1.5$ shown in Fig.~\ref{fig:viscosity_mie}\subref{subfig:LJ-12} for which $T_\mathrm{c} \approx 1.32$~\cite{thol2016}.
It is worth emphasizing that no fit parameters were used to achieve this accuracy.

For comparison, Fig.~\ref{fig:viscosity_mie}\subref{subfig:LJ-12} also includes the prediction for the shear viscosity using MET, as the Thol \textit{et al.} free energy expression~\cite{thol2016} makes it possible to apply MET to LJ fluids. At low density, the MET prediction is fairly accurate, similar to when $X_\mathrm{res} = \beta u_\mathrm{ex}$. At higher densities, MET tends to under-predict the shear viscosity and clearly performs worse. It should be noted though that the discrepancy is less severe at higher temperatures. Clearly $X_\mathrm{res} = \beta u_\mathrm{ex}$ captures more of the relevant physics underlying momentum exchange at elevated densities compared to $X_\mathrm{res} = X^\mathrm{MET}_\mathrm{res}$.

\begin{figure}
    \includegraphics[width=0.45\textwidth]{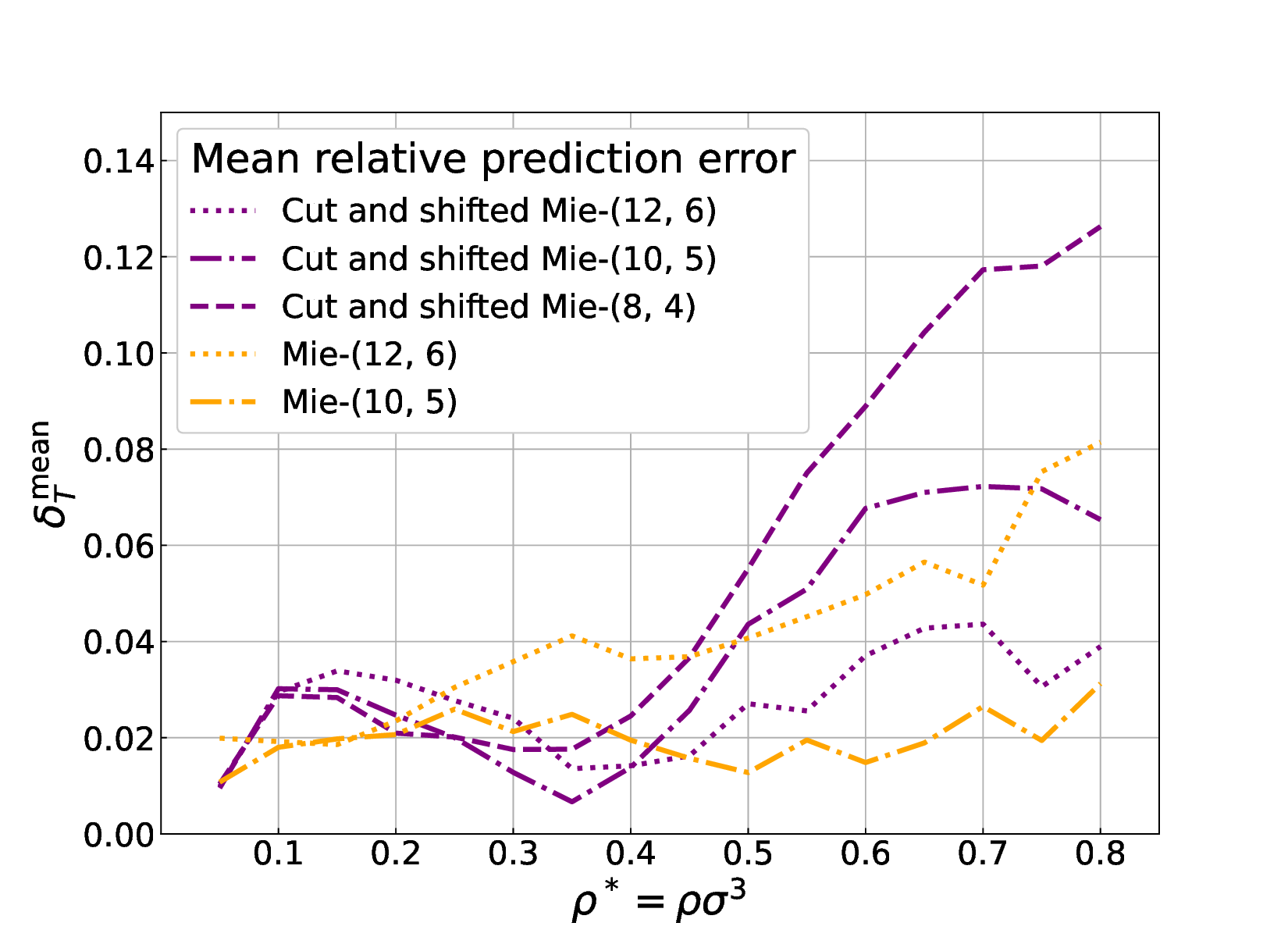}
    \caption{The relative prediction error $\delta^\mathrm{mean}_T$ averaged over temperature $T^*$ as a function of number density $\rho^*$. The errors are generally small when $\rho^* \lesssim 0.3$. Above this density particle cores are expected to continuously overlap, three-body correlations become significant, and the magnitude of the associated prediction error increases.}
    \label{fig:error_plot}
\end{figure}

Figure~\ref{fig:error_plot} shows how the mean relative prediction error averaged over temperature $\delta^\mathrm{mean}_T$ depends on the density for both the cut and shifted Mie fluids and full Mie fluids. For densities $\rho^* \lesssim 0.3$, the typical error remains low across all Mie fluids being investigated. When $\rho^* \gtrsim 0.35$, the errors generally get larger, although the magnitude of the errors depends on the exact interaction model and interaction parameters.
This is the density at which the particle cores are expected to continuously overlap and three-particle correlations become significant.
The effect is particularly noticeable for the cut and shifted Mie fluids (blue lines) where only the repulsive interaction is present.

In general, the high degree of correspondence between prediction and simulation data indicates that $X_\mathrm{res} = \beta u_\mathrm{ex}$ is well suited for predicting the shear viscosity of fluids characterized by a single Mie-type interaction.
It appears that $\beta u_\mathrm{ex}$ in the expression for $\hat{g}$ is an effective way to account for how the typical energy content of an interaction impacts momentum exchange. There is however no reason to believe that $X_\mathrm{res} = \beta u_\mathrm{ex}$ is always the best possible choice, or even the best choice for the systems under consideration in the present work. 
For future study, it would be interesting to investigate alternative expressions for $\hat{g}$ that rely on other residual properties for $X_\mathrm{res}$.

\subsection{The shear viscosity of diatomic LJ fluids}

\begin{figure}
    \centering
    \subfloat[]{
        \includegraphics[width=0.45\textwidth]{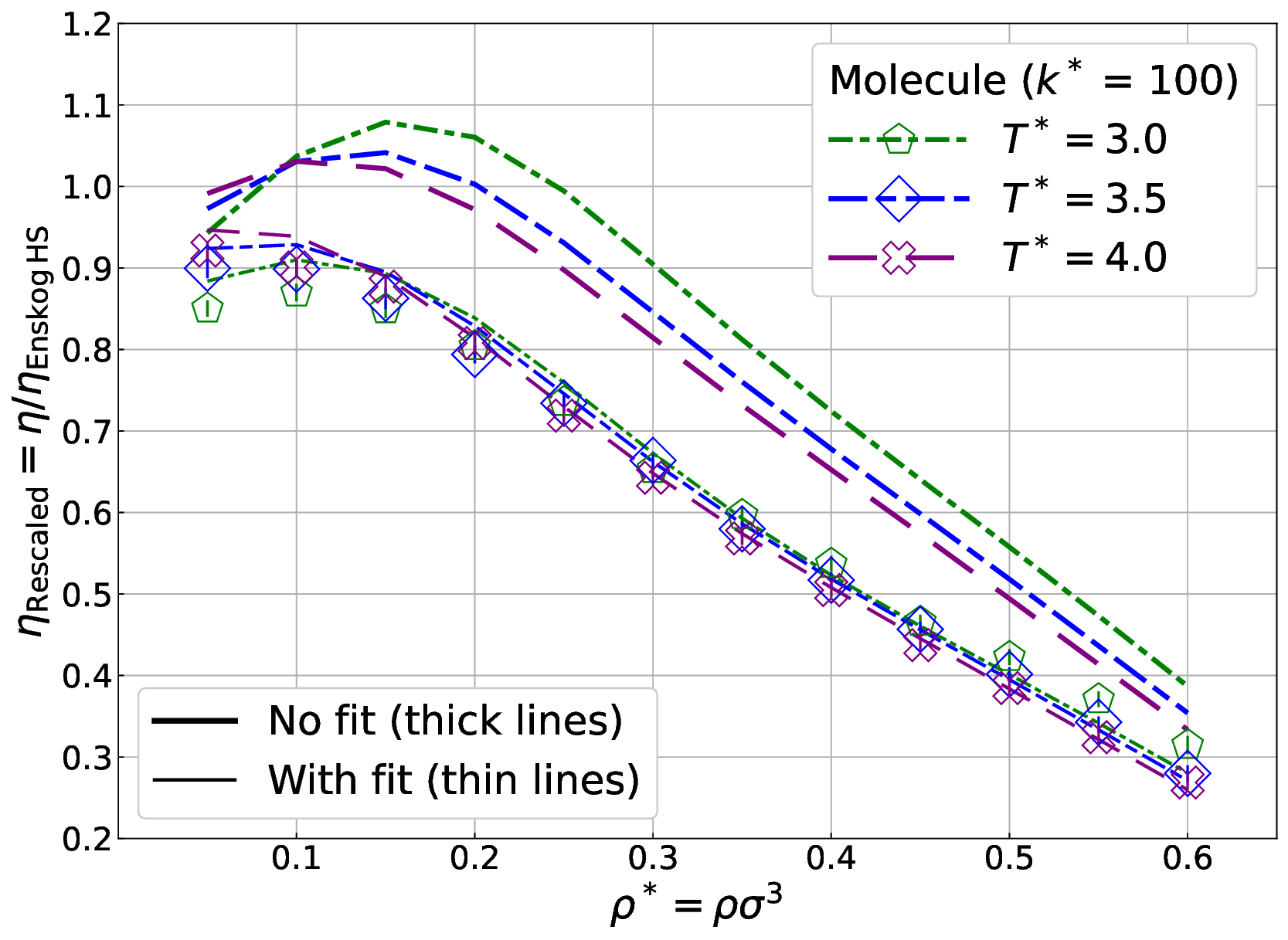}
        \label{subfig:Molecule-100}}
        \newline
    \subfloat[]{
        \includegraphics[width=0.45\textwidth]{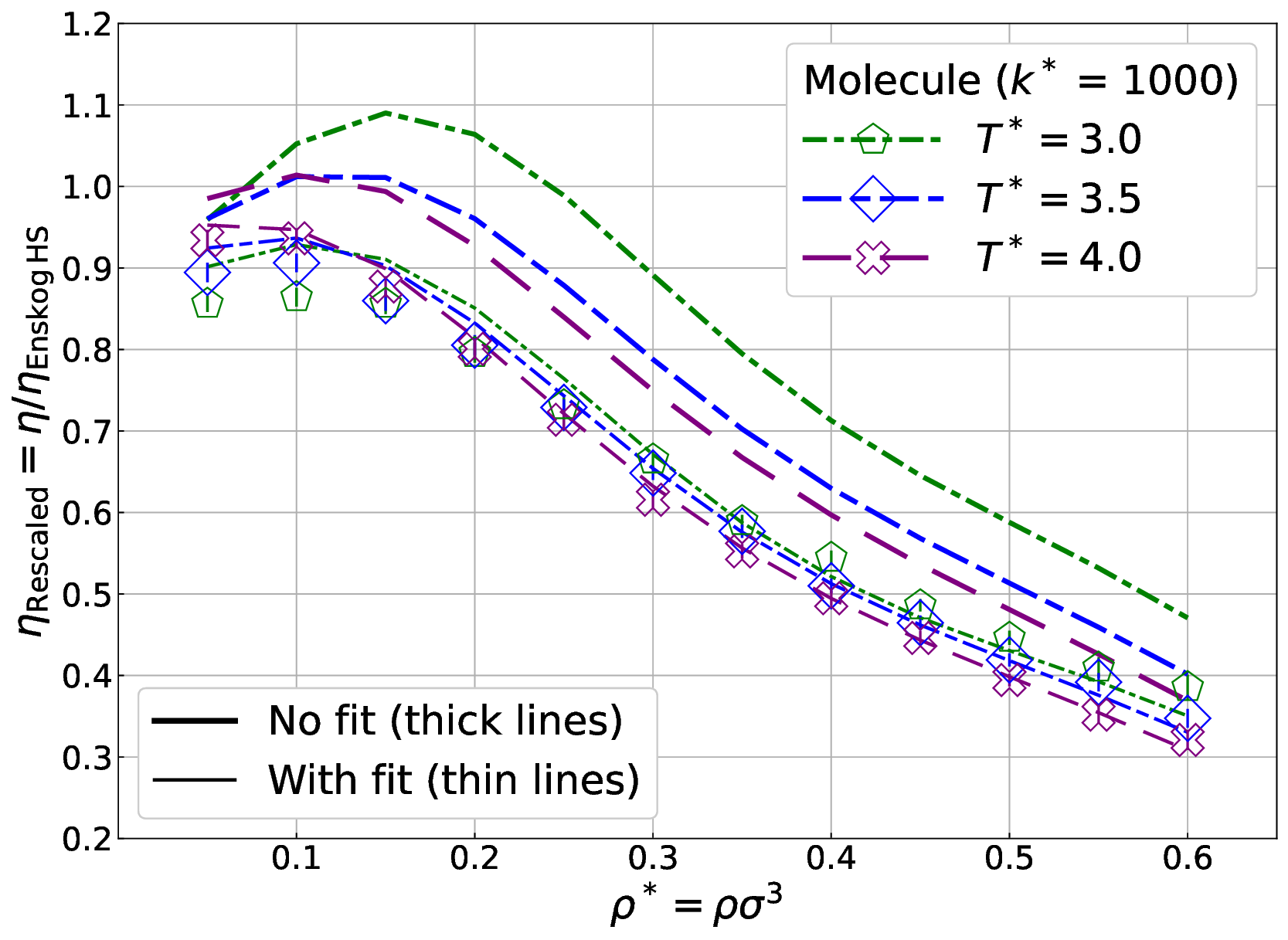}
        \label{subfig:Molecule-1000}}
    \caption{Rescaled shear viscosity from simulations and predictions for diatomic LJ fluids with spring constant $k^* = 100$ (a) and spring constant $k^* = 1000$ (b). Each temperature is assigned a symbol and associated dash style as denoted in the legend. The prediction lines are calculated using either $X_\mathrm{res} = \beta u_\mathrm{ex}$ (thick lines), or $X_\mathrm{res} = C\beta u_\mathrm{ex}$ (thin lines). In the latter case, $C$ is treated as a fit parameters. 
    The values used for $\eta_0$ are estimated using Eq.~(\ref{eq:estimating_eta_0_mean}), $\hat{g}$ is calculated based on equilibrium data from simulations as described in Sec.~\ref{sec:calc_hat_g}, and the low-density limit $\lim_{\rho \to 0} (\hat{g}/\rho)$ is estimated using Eq.~(\ref{eq:hat_g_lim_estimate}).
    The shear viscosity is rescaled using the Enskog prediction for a hard-sphere fluid with diameter $\sigma_\mathrm{HS} = (59/32)^{1/3}\sigma$ and mass $m = 2$.}
    \label{fig:viscosity_molecules}
\end{figure}

Figure~\ref{fig:viscosity_molecules} shows a comparison of the predicted viscosity with data for the shear viscosity of the simulated diatomic LJ fluids. All viscosity data and corresponding predictions presented in the figure have been rescaled using the Enskog prediction for hard spheres given by Eq.~(\ref{eq:Enskog}), setting $\sigma_\mathrm{HS} = (59/32)^{1/3}\sigma$ as this corresponds to the diameter of a hard sphere with excluded volume equal to that of a hard spherocylinder with width $1\sigma$ and total length $1.5\sigma$~\cite{Onsager1949}.

In the diatomic fluid, there are larger deviations between the theoretical prediction and numerical results than there were for the Mie fluids.
In both Fig.~\ref{fig:viscosity_molecules}\subref{subfig:Molecule-100} and Fig.~\ref{fig:viscosity_molecules}\subref{subfig:Molecule-1000} the prediction made using $X_\mathrm{res} = \beta u_\mathrm{ex}$ (dashed lines) is poor for any density $\rho^* > 0$, as the prediction immediately diverges from simulation data. Above $\rho^* \approx 0.2$ we found that the prediction error stabilizes, resulting in a prediction error between roughly $20\%$ and $30\%$. We can thus conclude that $X_\mathrm{res} = \beta u_\mathrm{ex}$ on its own is a poor choice for the diatomic LJ fluids. 

These deviations may be related to the fact that molecular interactions are no longer fully elastic, and thus momentum will be transferred differently.
We can try to account for this by adding a fit parameter $C$ that weighs the interaction term, i.e.,\
\begin{equation}
    \label{eq:hat_g-heuristic}
    \hat{g} = Z - 1 - C\beta u_\mathrm{ex}~.
\end{equation}
A similar approach is utilized by DHS kinetic theory which introduces an effective dipole moment thereby modifying the overall interaction strength. The resulting effective dipole moments were found to be considerably weaker than the actual dipole moments. Given this trend, one should expect $C < 1$ when fitting the resulting expression to data for inelastic model systems, which is in agreement with the values obtained for $C$ as reported in Table~\ref{tab:fit_params}, indicating that the effective interaction strength is weakened by the presence of inelastic interactions. 

The associated fitted lines are included in Fig.~\ref{fig:viscosity_molecules} (thin lines), and show clear improvement compared to $C=1$.
It cannot be ruled out that the proposed fit will absorb errors that are not associated with the inelastic nature of interactions. However, when the fit was applied to the perfectly elastic atomic Mie-type fluid models (not shown), it offered no significant improvement over the purely analytical approach, with the fitted $C$ being close to unity for all atomic Mie-type interaction models considered.

\begin{table}
\centering
\begin{tabular}{ p{0.2\linewidth} p{0.3\linewidth} p{0.2\linewidth} }
\hline
\hline
\vspace{1px} \\
\multicolumn{3}{c}{\textbf{Fit parameter $\boldsymbol{C}$ in Eq.~(\ref{eq:hat_g-heuristic}) }}\\
\vspace{1px} \\
\hline
 & $k^* = 100$ & $k^* = 1000$ \\
\hline
$T^* = 3.0$ & 0.875 & 0.885 \\
$T^* = 3.5$ & 0.881 & 0.913 \\
$T^* = 4.0$ & 0.873 & 0.911 \\
\hline
\hline
\end{tabular}
\caption{Fit parameter $C$ in Eq.~(\ref{eq:hat_g-heuristic}), when fitting Eq.~(\ref{eq:Enskog_modified}) to simulation data for the shear viscosity of diatomic LJ fluids.}
\label{tab:fit_params}
\end{table}

When $X_\mathrm{res} = C\beta u_\mathrm{ex}$ (thin lines), the fit is quite good for all temperatures considered and at most densities, both when $k^* = 100$ and when $k^* = 1000$. 
In both cases, $C$ remains similar for all temperatures investigated. It does however depend on the stiffness of the molecular bond.  This could be expected from the fact that this parameter should capture the inelasticity of the collisions.  In addition, as expected, it is closer to the elastic value of unity when the bond is stiffer. Depending on temperature, the mean relative error associated with these fits ranges from $\sim 1.5\%$ to $\sim 5\%$. These errors are of the same order as those associated with predictions of the shear viscosity of the atomic Mie-type fluids (Figs~\ref{fig:viscosity_mie_cut} and~\ref{fig:viscosity_mie}).

\subsection{Further discussion}

Hard-sphere approaches do have some advantages at low density. For example, Jervell and Wilhelmsen~\cite{Jervell2023} demonstrated a high level of predictive power in that density range.
However, hard-sphere approaches are fundamentally limited by the fact that an effective hard-sphere diameter will never be well suited for capturing the physics of momentum exchange at high density where particles are constantly partially overlapping. As such, effective hard-sphere approaches tend to work best at low density, well below the critical density of the fluid.

In contrast, although they require knowledge of an accurate free energy expression, approaches that make use of the general framework proposed in this paper should not be limited to low densities, as a well chosen residual function clearly contains information about the density dependence for the soft interactions that is missing from the pure hard-sphere description. The accuracy of our approach using $X_\mathrm{res} = \beta u_\mathrm{ex}$ at high densities, well above the critical density, is consistent with this.

%% file: Sections_new_structure/Conclusion.tex
\section{Conclusion}

We have derived a new framework for applying Chapman-Enskog theory to systems with soft and long-range interactions. Rather than relying on effective hard-sphere diameters, we define an exchange function that is expressed in terms of an appropriate thermodynamic relation. For hard spheres, this relation is proportional to the hard-sphere collision rate and depends only on the fluid compressibility factor $Z$. When applied to more complex fluids, the appropriate relation must include some additional residual contribution $X_\mathrm{res}$ that vanishes in the hard-sphere limit. We show that two existing extensions to Enskog theory, modified Enskog theory and kinetic dipolar hard-sphere (DHS) theory, both fit into this framework.

Inspired by kinetic DHS theory, we propose that the relation $X_\mathrm{res} = \beta u_\mathrm{ex}$ can be applied to a more general set of fluid models.
In order to test the relation, we predict the shear viscosity of three distinct simulated model systems. The first is a cut and shifted Mie potential, the second is the full Mie potential, and the third is a diatomic Lennard-Jones molecule.
Depending on the specific fluid model and interaction parameters, the fluids are simulated over a wide range of densities $0.05 \leq \rho^* \leq 0.8$ and temperatures $1.5 \leq T^* \leq 4.0$, where $\rho^*$ and $T^*$ denote the dimensionless Lennard-Jones density and temperature respectively.

When applied to the cut and shifted Mie fluids and the full Mie fluids,
the predictions closely followed the numerically calculated viscosity of the simulated fluids.
We find that predictive accuracy depends on the density.
At low and intermediate densities $\rho^* \leq 0.3$, the mean relative prediction error stays below $4\%$ in all cases. When $\rho^* > 0.35$, the associated mean relative errors get larger. 

When applied to diatomic molecular LJ fluids, the overall mean relative errors are larger, between $15\%$ and $30\%$. This is to be expected, as our proposed relation makes no attempt to account for the impact of microscopically inelastic collisions. To this end, we introduce a heuristic modification of $X_\mathrm{res}$, whereby $\beta u_\mathrm{ex}$ is scaled by a fit parameter $C$. 
The mean relative errors associated with the fit lie in the range of $1.5\%$ to $5\%$.

In general, we find that thermodynamic relations relying on $\beta u_\mathrm{ex}$ provide accurate descriptions of the shear viscosity across a wide set of fluid models and simulation parameters.
Further work is needed to understand how $X_\mathrm{res}$ generally depends on the underlying microscopic characteristics of the fluid.
Regardless, the success of our approach demonstrates that the limitations of Chapman-Enskog theory are not as severe as is often believed as long as the exchange function is chosen carefully.

%% file: Appendix/AppendixA.tex
\section{}
\label{app:first_order_term}

We here show, in detail, how to derive the first order term of the expansion for the residual function $X_\mathrm{res} = \beta u_\mathrm{ex}$. The derivation starts with the virial expansion for the compressibility factor $Z$, given by~\cite{HansenMcDonald2006}
\begin{equation}
    \label{eq:Z-expansion}
    Z - 1 = \sum_{n = 1}^\infty B_{n+1}(T)\rho^n~,
\end{equation}
where $B_{n+1}(T)$ are the virial coefficients. An expression for the dimensionless per-particle excess Helmholtz free energy $\beta f_\mathrm{ex}$ can be arrived at by making use of the thermodynamic relation $(Z - 1)/\rho = (\partial \beta f_\mathrm{ex} / \partial \rho)_T$. Substitution of Eq.~(\ref{eq:Z-expansion}) into the relation and subsequent integration yields
\begin{equation}
    \label{eq:beta_f-expansion}
    \beta f_\mathrm{ex} = \sum_{n = 1}^\infty \frac{1}{n} B_{n+1}(T)\rho^n~.
\end{equation}
By application of the thermodynamic relation given by $\beta u_\mathrm{ex} = -T ( \partial (\beta f_\mathrm{ex}) / \partial T )_{\rho}$, it follows that\break
\begin{equation}
    \label{eq:beta_u-expansion}
    \beta u_\mathrm{ex} = -\sum_{n = 1}^\infty \frac{1}{n} T \frac{d B_{n+1}(T)}{d T}\rho^n~.
\end{equation}
Finally, the second virial approximation, which truncates the virial series beyond $n = 1$, yields
\begin{equation}
    \label{eq:beta_u-low_density}
    \beta u_\mathrm{ex} \approx -T \frac{d B_{2}(T)}{d T}\rho~,
\end{equation}
in the low-density range which is what we set out to show. This is the exact same expression yielded by the second virial approximation of the MET residual function $X^\mathrm{MET}_\mathrm{res} = -T(\partial Z / \partial T)_\rho$ as given by Eq.~(\ref{eq:X_res_1}).

%% file: Appendix/AppendixB.tex
\section{}
\label{app:hat_g_exactness}

Here, we derive the expression for $X_\mathrm{res}$
that reproduces the expression $Z - 1 = V_\mathrm{excl} \rho g(\sigma_\mathrm{HS}^+) +  X_\mathrm{res}$
when applied to a system for which the pairwise interactions are given by combining the hard-sphere potential,
\begin{equation}
    \phi_\mathrm{HS}(r) = 
    \begin{cases}
        \infty, \quad & r < \sigma_\mathrm{HS} \\
        0,       \quad & r \geq \sigma_\mathrm{HS}
    \end{cases}~,
\end{equation}
with a long-range tail $\propto 1/r^n$.
We will show that $X_\mathrm{res} = (n/3)\beta u_\mathrm{ex}$ for such potentials.

We introduce the tail potential function $\phi_\mathrm{tail}(r) = a/r^n$, where $a$ and $n$ are arbitrary constants.
Insertion of $\phi(r) = \phi_\mathrm{HS}(r) + \phi_\mathrm{tail}(r)$ into Eq.~(\ref{eq:Z-integral}) yields
\begin{equation}
\label{eq:Z-integral-soft_tail}
\begin{split}
    Z - 1 =  - \frac{2}{3}\pi\rho\beta \int^\infty_0 dr g(r) r^3 \frac{d\phi_\mathrm{HS}(r)}{dr}
    \\
             - \frac{2 }{3}\pi\rho\beta \int^\infty_0 dr g(r) r^3 \frac{d\phi_\mathrm{tail}(r)}{dr}~.
\end{split}
\end{equation}
We denote the first term as $I_\mathrm{HS}$ and the second term as $I_\mathrm{tail}$.
The integral $I_\mathrm{HS}$ can be evaluated analytically by making use of the fact that $g(r) = y(r)\exp{[-\beta \phi(r)]}$~\cite{HansenMcDonald2006}, where $y(r)$ is the smooth cavity distribution function.  By
using the chain rule, i.e.\ $d\exp[-\beta \phi_\mathrm{HS}(r)]/dr = -\beta\exp[-\beta\phi_\mathrm{HS}(r)](d\phi_\mathrm{HS}(r)/dr)$,
we find that the hard-sphere term in Eq.~(\ref{eq:Z-integral-soft_tail}) can be rewritten as
\begin{equation}
\label{eq:Z-integral-soft_tail_1}
    \begin{split}
         I_\mathrm{HS} & = \frac{2}{3}\pi\rho \int^\infty_0 dr r^ 3 y(r) \exp{[-\beta \phi_\mathrm{tail}(r)]} \frac{d}{dr} \exp[-\beta \phi_\mathrm{HS}(r)] \\
                     & = \frac{2}{3}\pi\sigma_\mathrm{HS}^3 \rho y(\sigma_\mathrm{HS})\exp{[-\beta \phi_\mathrm{tail}(\sigma_\mathrm{HS})]}~.
    \end{split}
\end{equation}
So far, this derivation is very similar to the one given by, among others, Ref.~\cite{HansenMcDonald2006} for pure hard sphere systems that ultimately leads to Eq.~(\ref{eq:Z-HS}).
Finally, Eq.~(\ref{eq:Z-integral-soft_tail_1}) can be further simplified by noting that in the limit $r \to \sigma_\mathrm{HS}^+$, $g(\sigma_\mathrm{HS}^+) = y(\sigma_\mathrm{HS})\exp{[-\beta \phi_\mathrm{tail}(\sigma_\mathrm{HS})]}$ and so 
\begin{equation}
\label{eq:Z-integral-soft_tail_1_final}
        I_\mathrm{HS}= \frac{2}{3}\pi\sigma_\mathrm{HS}^3 \rho g(\sigma_\mathrm{HS}^+) = V_\mathrm{exc}\rho g(\sigma_\mathrm{HS}^+)~,
\end{equation}
where we have used that $V_\mathrm{exc} = 2\pi\sigma_\mathrm{HS}^3/3$.

The second term in Eq.~(\ref{eq:Z-integral-soft_tail}), $I_\mathrm{tail}$, can be simplified by noting that $d\phi_\mathrm{tail}(r)/dr = -(n/r) \phi_\mathrm{tail}(r)$, so that
\begin{equation}
\label{eq:Z-integral-soft_tail_2}
    \begin{aligned}
    I_\mathrm{tail} = \frac{2n}{3} \pi\rho\beta \int^\infty_0 dr g(r)r^2 \phi_\mathrm{tail}(r) = \frac{n}{3} \beta u_\mathrm{ex} ~,
    \end{aligned}
\end{equation}
where the second step follows by comparing Eq.~(\ref{eq:u_ex-integral}), and noting that only $\phi_\mathrm{tail}(r)$ contributes to to the total excess internal energy. 

Insertion of both Eq.~(\ref{eq:Z-integral-soft_tail_1_final}) and Eq.~(\ref{eq:Z-integral-soft_tail_2}) into Eq.~(\ref{eq:Z-integral-soft_tail}) gives
\begin{equation}
\label{eq:hat_g_exact}
    Z - 1 =  V_\mathrm{excl} \rho g(\sigma_\mathrm{HS}^+) + \frac{n}{3} \beta u_\mathrm{ex}~.
\end{equation} 
which is what we set out to show.
In the special case when $n=3$, $X_\mathrm{res} = \beta u_\mathrm{ex}$.